\documentclass[ag]{copernicus2}

\begin{document}

\title{Solar wind magnetic turbulence: Inferences from spectral shape 
}

\author[1]{R. A. Treumann}
\author[2]{W. Baumjohann}
\author[2]{Y. Narita}

\affil[1]{Department of Geophysics and Environmental Sciences, Munich University, Munich, Germany}
\affil[2]{Space Research Institute, Austrian Academy of Sciences, Graz, Austria}

\runningtitle{Magnetic turbulence in solar wind}

\runningauthor{R. A. Treumann, W. Baumjohann and Y. Narita}

\correspondence{R. A.Treumann\\ (rudolf.treumann@geophysik.uni-muenchen.de)}




\maketitle

Some differences between theoretical, numerical and observational determinations of spectral slopes of solar wind turbulence are interpreted in the thermodynamical sense. Confirmations of turbulent Kolmogorov slopes in solar wind magnetic turbulence and magnetohydrodynamic simulations exhibit tiny differences. These are used to infer about entropy generation in the turbulent cascade and to infer about the anomalous turbulent collision frequency in the dissipative range as well as the average energy input in solar wind turbulence. Anomalous turbulent collision frequencies are obtained of the order of $\nu_{an}\approx 200$ Hz. The corresponding stationary solar wind magnetic energy input into magnetic turbulence in the Kolmogorov inertial range is obtained to be of the order of $\approx 50$ eV/s. Its thermal fate is discussed.

\section{Introduction}
\label{intro}
Turbulence in plasmas whether or not magnetic --  like in the solar wind and the magnetospheric tail -- is known to exhibit a number of spectral ranges. According to the Kolmogorov theory of stationary turbulence, turbulent energy is continuously injected at large scales $\ell< L$ shorter than the linear dimension $L$ of the volume under consideration. It causes the evolution of large-scale eddies at small wave numbers $k_\mathit{in}=2\pi/\ell >2\pi/L$. Nonlinear cascading brings shorter scales into play. In collisionless systems the initiation of the cascading is mainly due to the production of harmonic sidebands. Spectral washing out and gap filling between the sidebands is produced by still badly understood higher nonlinearities like collisionless spectral broadening of the sidebands. Since injection is at small $k$, the first group of sidebands lie close together in wavenumbers. Broadening becomes important at large $k$. Cascading implies energy distribution from the initial large-scale eddies at time $t= -\infty$ to a multitude of shorter-scale smaller eddies of increasing wave number $k>k_\mathit{in}$. Ultimately it forms a quasi-continuous spectrum which possesses some $k$-intervals of a characteristic power law slope. According to \citet{kolmogorov1941a,kolmogorov1941b} the power-law range of the spectrum is the inertial range.

Observations in solar wind magnetic turbulence and theory have invested much effort into the determination of the shape and formation of the inertial range spectrum. They have settled on the confirmation of the inertial Kolmogorov magnetic-power spectrum $W(k)\propto k^{-3/2}$, recently verified based on spacecraft data \citep{alexandrova2009,brown2015}. Detection of anisotropy in the power spectra of magnetic fluctuations showed the expected and well understood differences between parallel and perpendicular spectra \citep{wicks2012,horbury2012}. 

Nonetheless there are a number of unsettled problems concerning the extension and shape of the inertial range. It is intuitive that once the eddy-scale becomes short enough, such that it matches any relevant dissipation scale at wavenumber $k_d$, the spectrum will be cut-off and the turbulent energy will be dissipated. There has been much discussion where this happens and which process is the relevant one \citep{alexandrova2009,sahraoui2009,sahraoui2013}. Discussion is also going on in which way the cascade is formed and maintained throughout the inertial range as this process has by far not been understood yet. Kolmogorov's scaling theory \citep{kolmogorov1941a,kolmogorov1941b,kolmogorov1962} does not provide a mechanism of turbulence. Claims have been put forward that chaotic intermittency would be responsible in generating the cascade and that non-Gaussian statistics would apply. This is all possible but still does not say much about the underlying physics, not providing physical insight into the process of turbulence, i.e. into the kind, number and interaction of the waves that are involved into the cascade. 

Observations \citep{perri2011,perri2012} suggest that the turbulence in the ion inertial range, on scales less than the thermal ion gyroradius, evolves into current vortices. This is important in view of the fact that in this domain non-magnetic ions practically drop out of the dynamics that affects the evolution of magnetic fluctuations. Magnetic transport and fluctuations are exclusively carried by the electron component here. The turbulent vortices and eddies become electron vortices, and the turbulent current filaments are almost pure electron current fuilaments. Unrestricted by any other process, this cascading proceeds to ever shorter scales $\ell$ and larger wave-numbers $k$. 

Whence the electron-vortex-mediated cascade ultimately reaches into the dissipation range at scales $\ell\sim\ell_d$ and $k\sim k_d=2\pi/\ell_d$, the energy is converted into heat or into some other non-turbulent form of energy and is extracted from turbulence. At this point the cascade terminates and the spectrum cuts off, mostly believed in an exponential or modified-exponential way $W(k> k_d)\propto g(k)\exp (-\eta k^{\zeta})$ with $\eta$ some typical dissipation scale, $\zeta$ some exponent, and $g(k)$ generally some polynomial. Observations of magnetic turbulence spectra in the solar wind \citep{goldstein1995,alexandrova2009,sahraoui2009,sahraoui2013,perri2012} have confirmed various dissipative spectral types of different shape at large $k$ including exponents in the interval $\frac{1}{2}\lesssim\zeta\lesssim 2$. This non-power law $k\gtrsim k_d$-range, whose existence was proposed originally by \citet{kolmogorov1941a}, has been interpreted as the turbulent dissipation range. 

The dissipation on the electron scales is then solely provided by the electron dynamics. Since this is restricted to the short scale current filaments of either electron gyroradius $r_{ce}$ or electron-inertial scale $\lambda_e=c/\omega_e$ it is most probably due to spontaneous collisionless reconnection \citep{treumann2015a,treumann2015b}. Without specification of the detailed mechanism, this was suggested implicitly already \citep{sundkvist2007,retino2007,sahraoui2009,perri2011} who noted that in the solar wind turbulent dissipation should be located most probably in electron-scale currents.

This classical picture raises two questions. The first is the question on the cascading mechanism under stationary conditions of turbulence. The second is the question, how far down in scale $\ell$ or up in wave number $k =2\pi/\ell$ the turbulent spectrum can, in principle, extend if a certain amount $\epsilon=\Delta E_{in}/m \Delta t$ of energy per mass $m$ is temporarily injected for the time $\Delta t$ and non-dissipatively distributed over an increasing number of smaller-scale eddies where the larger part of the injected energy is stored. Does it at all reach the non-collisional dissipative range and if to what amount? And for how long must the injection of energy be maintained for this to happen? Here, we are going to investigate only Kolmogorov's question on the shape of the stationary spectrum at stationary injection rate $\epsilon=$ const.

\section{The stationary inertial range spectrum}
It seems to be clear that under ideal conditions of stationarity, with constant injection of a fixed amount of energy per time and a constant dissipation efficiency in the dissipative range, the spectrum will necessarily extend into the dissipative range. This is suggested by the Kolmogorov theory. Energy must flow down the spectrum into the dissipation range where it is extracted from turbulence and disappears in some form which does not belong to turbulence anymore and does not affect its shape and evolution. The dissipated energy will then be just the amount of energy that exceeds the energy that, initially, has been necessary to create the eddies that form the spectrum and extend the spectrum into the dissipation range while maintaining its stationary, not necessarily power-law slope. 

Under real conditions the eddies will consume some fraction of the injected energy in order to feed themselves and maintain their existence. Otherwise they would resemble ideal reversible Carnot machines. This cannot hold true for the simple reason that any eddy in order to generate the next smaller-scale eddy which extends the cascade must do some work on this eddy. Such processes are not taken care of in Kolmogorov theory which assumes that the injected energy supply at the injection scale is constant and conserved throughout the entire spectral decay until it arrives at the dissipation range. The same assumption also underlies Kolmogorov's followers \citep{heisenberg1948,obukhov1962}.  

In the absence of any convincing theory of the formation of the inertial range spectrum one seeks confirmation by numerical experiments.  Very high resolution numerical simulations of magnetohydrodynamic turbulence have only become possible in recent years after the availability of supercomputers. Such series of high-resolution simulations of MHD-turbulence have been preformed by \citet{bereznyak2011,bereznyak2014} with the goal to verify the spectral slope in the inertial range generally predicted by Kolmogorov. 

 \citet{bereznyak2014} essentially achieved his goal. His simulations confirmed Kolmogorov's prediction in application to MHD turbulence. In addition he identified a small though significant deviation  from the nominal Kolmogorov spectral slope in the inferred inertial spectral range. This inferred slope difference amounts to $\Delta\alpha\approx 1.70-5/3\approx 0.033$ \citep{bereznyak2011,bereznyak2014}. Since it is tiny, it has been taken as a nearly perfect identification of Kolmogorov's prediction and the complete validity of the theory. Indeed, the confirmation of the theory by the simulations is reasonable and as perfect as one could expect. It puts other claims of either much flatter $\sim k^{-1}$ or substantially steeper spectra $\sim k^{-2}$ into serious doubt -- at least what concerns stationary MHD turbulence. \citet{alexandrova2009} also reported an observation based slope of $\sim k^{-1.7}$ which, however, is approximate and taken as confirmation of Kolmogorov.

One should nevertheless keep in mind that the slope of a power law spectrum is a very sensitive function of the underlying physics. Subtle changes in slope may imply violently different physical processes. The above $\sim$\,2\% deviation identified in the high-resolution simulations \citep{bereznyak2014} is by no means small enough to warrant its neglect. Any observations in space, in particular in the solar wind \citep{goldstein1995,zhou2004,alexandrova2009}, which also did confirm Kolmogorov's theory were, however, much less precise than the results of the simulations. It might be argued that the simulations have been performed in the strong magnetic field limit. This argument is, however, not valid in view of the solar wind observations. Field strength in MHD plays a role only in causing anisotropy as observed in the above cited work \citep{horbury2012,wicks2012}.

Observationally detecting a deviation in slope so subtle as the above one has so far been completely illusive. Stating this conjecture, the  confirmation of the Kolmogorov theory in application to MHD turbulence is by no means in doubt. The deviation in slope  is, however,  large enough to argue that it reflects some interesting process that is not covered by the Kolmogorov theory or its followers. It tells that consumption of some small fraction of the injected energy is required for the maintenance of the stationary spectrum of the turbulence. 

\section{Physical content of a slope mismatch}  
Kolmogorov's theory of the shape of the power spectrum $W(k)$ of stationary, i.e. time-independent turbulence can be reduced to a simple dimensional analysis, assuming that the inertial spectral energy density per unit mass $W(k)\sim W/k$ at a particular wavenumber $k$ locally depends only on $k$ and the constant energy inflow rate $\epsilon \sim W/t$. Here $W$ is the energy per mass. This implies that dimensionally 
\begin{equation}
W(k) = C_K \epsilon^\alpha k^\beta \quad \longrightarrow\quad l^3t^{-2}=(l^2t^{-3})^\alpha l^{-\beta}
\end{equation}
which yields Kolmogorov's result $\alpha=\frac{2}{3},\, \beta=-\frac{5}{3}$ and holds for $\epsilon$ constant, i.e. independent of wavenumber $k$. Here $C_K$ is a normalisation constant, in fact Kolmogorov's constant of value $1.6<C_K<1.7$. A collection of these relations is found in many places, for instance in \citet{biskamp2003}. We repeat that no dynamical theory exists for this scaling even though Kolmogorov himself and all followers have provided plausibility arguments. Any those arguments ultimately reduce to the above scaling.  

A similar scaling holds for MHD turbulence, because MHD waves (either Alfv\'en or magnetosonic) have linear dispersion $\omega=V_\mathit{mhd}k\sim V_Ak$ and thus, like in the assumption on the eddy-crossing time underlying Kolmogorov's fluid theory, are linear in wavenumber $k$. The typical eddy-crossing or ``turn-over'' time is of the order of the  Alfv\'en time $t_\mathit{mhd}\propto l/V_A$. 

In the ion-inertial range, on the other hand, kinetic Alfv\'en waves come into play to replace ordinary MHD waves. This should move the crossing time over to $t_\mathit{kA}\propto l^2$, which yields a somewhat steeper spectrum with $\beta_\mathit{iir}={-7/3}$. It is a bit surprising that this slope has barely ever been confirmed, neither in observations nor in simulations including the ion-inertial range. The reason can be found when inspecting the linear dispersion relation of kinetic Alfv\'en waves $\omega=k_\|V_A(1+\tilde{r}_{ci}^2k_\perp^2)^{1/2}$, with $\tilde{r}_{ci}$ a temperature modified ion-gyroradius. The second term in the brackets is a non-negligible correction which, however yields a $k_\|k_\perp$ dependence with $k_\|\ll k_\perp$. Any quadratic wavenumber dependence thus becomes weakened. Though kinetic Alfv\'en waves dominate the ion-inertial range because its spatial extension is of the order of the ion-inertial length, which is the perpendicular scale of these waves, they seem to have no violent effect on the turbulence. Waves with $k^2$-scaling are whistlers. Their propagation is mostly parallel, however. 

The postulated independence of the energy inflow rate on wave number, i.e. its constancy over the entire inertial range in stationary turbulence, may be doubted for thermodynamic reasons. One should expect that some energy would be retained at any wavenumber $k$ for the maintenance of the particular eddy even in the case that there is no real viscous or resistive dissipation present. If this would not be the case, the various eddies would, as already noted, represent ideal \emph{reversible} Carnot machines. This, in a real physical system, will hardly be realised  because of the thermodynamic argument that the eddy is an open system in contact with other systems. The mere completely lossless transport of energy contradicts thermodynamics. 

Indeed, comparison with the simulations \citep{bereznyak2014} suggests that the energy flux  cannot be constant across the spectrum. It should become itself a function of the wave number. In order to reproduce the slope this dependence will be weak :
\begin{equation}
\epsilon(k)=\epsilon_\mathit{in}(k/k_\mathit{in})^{-\delta}
\end{equation}
with $\delta=1/20$ read from the simulations, with $\epsilon_\mathit{in}$ energy injection rate at injection wavenumber $k_\mathit{in}$. Here $k_\mathit{in}$ is the wavenumber at the low-$k$ end of the inertial range cascade.

With reference to this wavenumber dependence of the energy flux, the power in the Kolmogorov law becomes $-\beta\approx 5/3+2\delta/3$. This yields the scaling
\begin{equation}
W(k)= C_K \epsilon_\mathit{in}^{2/3}k_\mathit{in}^{2\delta/3}k^{-5/3-2\delta/3}
\end{equation}
which is the simulated slightly steeper spectrum than the Kolmogorov spectrum. It accounts for the energy loss in the energy transfer across the eddy which forms at wavenumber $k$. Using the above inferred value of $\delta=1/20$, we thus have $\beta=51/30$ and
\begin{equation}
W(k)= C_K \epsilon_\mathit{in}^{2/3}k_\mathit{in}^{1/30}k^{-51/30}
\end{equation}

One may note that this is different from any proposal for a modification of the Kolmogorov spectrum  based on some nonlinear approximation, attempts that had been put forward by \citet{heisenberg1948} and by \cite{obukhov1962}. 

At any fixed $k$ the difference between the simulated spectrum and the Kolmogorov spectrum then gives an estimate of the spectral energy that at wave number $k$ is retained in the local eddies and is not anymore accessible to the turbulence: 
\begin{equation}\label{eq-wk}
\delta W(k)= C_K\epsilon_\mathit{in}^{2/3}k^{-5/3}[1-(k_\mathit{in}/k)^{2\delta/3}] 
\end{equation}
This expression vanishes of course at $k=k_\mathit{in}$ and maximises at $k_m=1.8\,k_\mathit{in}$. The maximum difference becomes $\delta W_m(k_m)\approx 0.01\,C_K\epsilon^{2/3}_\mathit{in}k^{-5/3}_\mathit{in}$. Integration gives the energy that for an injected energy flux $\epsilon_\mathit{in}$ is consumed in the inertial range eddies in order to keep them going up to $k$:
\begin{equation}
\Delta W = \int_{k_\mathit{in}}^k \delta W(k)\,\mathrm{d}k
\end{equation}
This is the amount
\begin{eqnarray}
\Delta W(k)&=& \frac{3}{42}C_K\epsilon_\mathit{in}^{2/3}k_\mathit{in}^{-2/3}\times \\
&\times&\left\{ 1-\frac{3}{2}\left(\frac{k}{k_\mathit{in}}\right)^{-\frac{2}{3}}\left[1-\left(\frac{k_\mathit{in}}{k}\right)^\frac{1}{30}\right]\right\}
\end{eqnarray}
The second term in the braces is close to $\sim 0.03$ as can be checked by assuming that the total inertial range covers just one order of magnitude. Simulations as well as observations in the solar wind show that this is marginally the case. Most inertial ranges are somewhat though not much longer. Hence the total consumed energy up to a given $k$  is approximately given by the coefficient of the expression in the braces on the right with only a small $k$-dependent correction. Since the correction is weak, the total energy supply to the inertial range can be well approximated by 
\begin{equation}
\langle\Delta W\rangle \approx \textstyle{\frac{1}{2}}k_m\delta W(k_m)
\end{equation}
taken in the inertial range $\Delta k= k_d-k_{in}$, with $k_d$ the wavenumber of dissipation.

\section{Comparison to solar wind magnetic turbulence}
Before proceeding, it is of interest to compare the above spectrum with actual measurements of solar wind turbulence \citep{brown2015}.

\begin{figure}
\centerline{\includegraphics[width=0.5\textwidth]{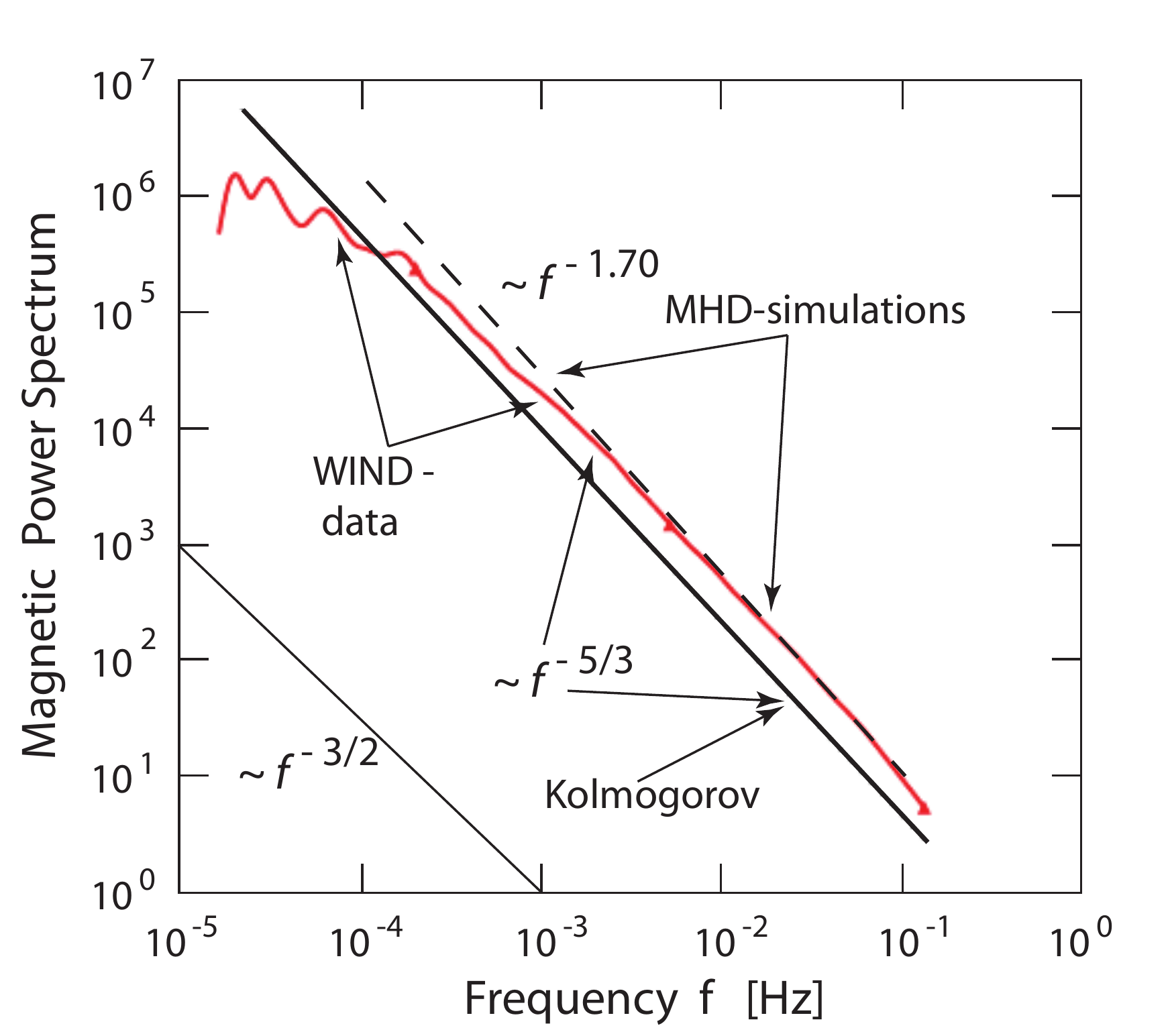}}
\caption{Spectrum of magnetic turbulence in the solar wind as obtained from an 8-day continuous observation \citep[spectral data taken from][]{brown2015}. The ordinate is in arbitrary powers of magnitude. The red curve are the spectral data calculated by the above authors in the time frequency domain under the assumption of validity of the Taylor-hypothesis which, for the present data set, has been found to be applicable. The black line shows the Kolmogorov spectrum. The authors interpret this spectrum as being a confirmation of the Kolmogorov spectrum which, in the light of our discussion, is reasonable. However, closer inspection shows that the power law domain between $0.01<f<0.1$ Hz is fitted even better by a power law of slope $\beta=-1.70$ (dashed line) which is in perfect agreement with the MHD-simulation result of \citet{bereznyak2014}. This strictly log-log linear inertial power-law range follows a $\lesssim 1$ order long approximate Kolmogorov-inertial range and covers $\sim$\,1 order of magnitude. One may note that the slope of the low-frequency part of the spectrum before entering the Kolmogorov range is almost exactly -3/2, which is the Iroshnikov-Kraichnan slope shown in the lower-left corner of the figure. It thus indicates that the turbulence in this domain is of about two-dimensional Iroshnikov-Kraichnan character.}\label{fig-1}
\end{figure}

Figure \ref{fig-1} shows the magnetic power spectrum of the eight-day long continuous Wind spacecraft observation of magnetic turbulence in the solar wind by \citet{brown2015}. Plotted is the power spectral density $\delta B^2(f)$ in arbitrary units as function of frequency $f$. It is assumed that the Taylor hypothesis is applicable which tells that any spatial structures are convected at same constant speed and, therefore, the Doppler-shifted frequency spectrum reflects the spatial spectrum of the turbulence. \citet{brown2015} have carefully checked and confirmed its validity by determination of the correlation times in their frequency range and for the high solar wind convection velocities of $V_{sw}\sim 600$ km/s during their observations. 

The red curve in Fig.\,\ref{fig-1}  is the magnetic spectrum in logarithmic scaling in arbitrary units. Following an injection phase at frequencies $f<5\times10^{-4}$ Hz the spectrum evolves into a power law which persists roughly 2 decades in frequency and is taken as an almost perfect Kolmogorov spectrum when compared to the $f^{-5/3}$ reference line. The uncertainty in the observations is of the order of the thickness of the red line and is thus negligible. 

Closer inspection reveals two interesting properties of the spectrum. The first is that at frequencies $5\times10^{-4}< f<2\times10^{-3}$ Hz the spectrum exhibits an almost exact Iroshnikov-Kraichnan slope of power $-3/2$ which \citet{brown2015} ignore. This observation confirms the claim put forward by \citet{treumann2015a} on the composition of solar wind spectra of magnetic turbulence. This large-scale range corresponds to practically two-dimensional turbulence. It extends rougly up to $\sim 3\times10^{-3}$ Hz in frequency. At the observed Alfv\'en speed of $V_A\approx 60 $\,km/s this corresponds to scales of $\ell\sim 2\times10^4$ km. We note in passing that the large-$k$ termination of this range cannot be brought in agreement, however, with neither correspondence with the ion inertial scale $\lambda_i=c/\omega_i$ which would be of the order of $\lambda_i\sim 650$\, km, nor the ion gyroscale as the latter would require unreasonably high ion temperatures. 

The second property of the spectrum is that, in the higher frequency inertial range, it can be approximated substantially better by a spectral slope $f^{-1.70}$ than with the Kolmogorov slope. This is shown as the thin dashed line. Its fit is almost perfect in a slightly shorter range of $\sim 1$ decade in frequency in the interval $10^{-2}< f< 10^{-1}$ Hz.

This fit agrees with the simulation results of \citet{bereznyak2014} and also with the less precise observations of \citet{alexandrova2009}. It is the transition region between the Iroshnikov-Kraichnan spectrum and the Bereznyak-spectrum where the slope of the spectrum becomes perfectly Kolmogorov within a range of about one decade in frequency. These fits confirm Kolmogorov's theory while they demonstrate the relevance of the difference in slope. 

In order to apply the estimate of the amount of energy that is consumed in the Kolmogorov-Bereznyak inertial range, the Wind data must be reinterpreted. The spectrum in Fig.\,\ref{fig-1} is the magnetic spectral energy density
\begin{equation}
[\delta B(\omega)]^2 = (\Delta t)^{-1}\left|\int_0^{\Delta t} \mathrm{d}t\,\delta B(t)\,\exp\,(-i\omega t)\right|^2
\end{equation}
and $\Delta t =8$ days is the total observation time. This implies that 
\begin{equation}
[\delta B(\omega)]^2 \mathrm{d}\omega = \delta W(k) \mathrm{d}k
\end{equation}
with d$\omega/$d$k\approx V_A$. This then yields
\begin{equation}
\langle\Delta W_B\rangle \approx\textstyle{\frac{1}{2}}f_m [\delta B(f_m)]^2 = \textstyle{\frac{1}{2}}\delta W(k_m)V_A^{-1}
\end{equation}
Here $\langle\Delta W_B\rangle =\langle\Delta{B^2}\rangle$. Magnetic fluctuations at the higher frequencies are of the order of $1\ \mathrm{nT}\,\lesssim|\,\delta B|\,\lesssim 2$ nT. With the measured plasma density $N\sim 10^7$ m$^{-3}$, the turbulent magnetic energy per particle in this  frequency range is $W_B(f)\approx 1$ eV which is $\lesssim10$\% of the magnetic energy per particle in the solar wind. This can be compared to the electron temperature of $T_e\sim 100$ eV. With these values we find 
\begin{equation}\label{wb}
\langle\Delta W_B\rangle \approx 4\times10^{-3}W_B
\end{equation}
As expected, this $4\%_{{0}}$ per mille average energy attribution per particle in the Kolmogorov range to the production of entropy is very small.  It is nevertheless physically significant as it adjusts the Kolmogorov theory to the thermodynamic requirements. In this estimate we used the Alfv\'en speed. If we would have referred to the Taylor hypothesis on which the generation of the spectrum has been based \citep{brown2015} then $V_A$ should be exchanged with $V_{sw}\approx 600$ km/s. This reduces the energy loss $\langle\Delta W_B\rangle$ by another factor of 10.

\section{Inferences on the dissipation rate} 
The solar wind is an extended medium hosting magnetic turbulence that, on the larger scales, seems to be homogeneous and stationary. It provides information on magnetic turbulence in the unaccessible solar corona where the solar wind magnetic fluctuations are generated.  The aim is the identification of the mechanism by which magnetic turbulence in the solar wind is destroyed on the shortest scales. Various different processes have been put forward, depending on the observation of the different dissipative decays of the measured spectra. 

It is agreed today that the main collisionless dissipation occurs on the electron scale at or below the electron gyroradius, which implies the extension of the spectral range down to those scales. Since, however, electrons become non-magnetic here and any \emph{magnetic} turbulence on those scales necessarily evolves into the generation of electron-scale current filaments and vortices, it becomes gradually clear that under the collisionless conditions on these scales \emph{spontaneous reconnection} is the main mechanism of dissipation \citep{treumann2015a,treumann2015b}. Electron-scale currents were indeed observed  by \citet{retino2007} in the magnetosheath, and by \citet{perri2011} in the solar wind. 

We note that the termination of the frequency-transformed \citet{brown2015} spectrum at frequency $f\lesssim0.1$\,Hz corresponds to a wavenumbers $k\lesssim \omega/V_{sw}\approx 0.6/600=10^{-3}\ \mathrm{km}^{-1}$ or wavelengths the order of $\lambda\gtrsim 6000$\,km. This is substantially larger than the respective electron-inertial or electron-gyro scales such that no reconnection-caused dissipation is expected to affect the spectrum at these scales. This is in excellent agreement with the validity of the Taylor hypothesis in the observational range of \citet{brown2015}.

In the ion inertial range the magnetic turbulence is due to electron dynamics, and the eddy-turnover Alfv\'en velocity must be replaced by the electron-Alfv\'en speed $V_{Ae}\approx \sqrt{m_i/m_e}V_A$, which is of the order of $V_{Ae} \gtrsim 10^3$ km\,s$^{-1}$. Whichever kind of waves the electron-turbulence is built of -- whistlers, electromagnetic electron-drift waves etc. -- this velocity exceeds the solar wind speed. Starting from the ion-inertial range the Taylor hypotheses may become violated towards larger $k$. Eddies that propagate downstream with the solar wind will shift wavelengths $\lambda<\lambda_i$ and $\lambda<r_{ci}$ into the high-frequency range and extend the inertial range frequency spectrum towards high frequencies. Upstream eddies instead contribute to lower frequencies and flatten the inertial range spectrum. Such effects may affect the higher frequency dissipation range. It is clear that in this case the identification of the frequency corresponding to the transition into the inertial range becomes problematic, as any spectral breaks in the slope will be smeared out by this Doppler-effect.

Because of the smallness of the amount of energy attributed to keeping the cascade going, we may adopt Kolmogorov's result on the dissipation scale and neglect the weak dependence of the energy flux on wavenumber. Then we have for the dissipation scale
\begin{equation}
k_d\approx 2\pi\left(\epsilon_\mathit{in}/D_\nu^3\right)^{1/4}
\end{equation}
where $D_\nu=(\mu_0\sigma)^{-1}$ is the magnetic diffusivity caused by the spontaneous reconnection in the dissipation range with $\sigma=\epsilon_0\omega_e^2/\nu_\mathit{an}$ the conductivity with respect to any effective anomalous resistivity or collision frequency $\nu_\mathit{an}$. 

Let us assume that the turbulent dissipation indeed results from spontaneous reconnection in a multitude of collisionless turbulent electron-scale current filaments which fill the turbulent volume in the range where the electron-cyclotron radius exceeds the electron-inertial length $r_{ce}\geq\lambda_e=c/\omega_e$. We may then assume that either $k_d\sim 2\pi/\lambda_e$ or $k_d\sim 2\pi v_e/\omega_{ce}$. 

If we use the electron-inertial length, then  the Kolmogorov dissipation scale yields for the anomalous collision frequency caused by spontaneous reconnection
\begin{equation}
\nu_\mathit{an}\sim \left(\epsilon_\mathit{in}/\lambda_e^2\right)^{1/3}\quad \mathrm{at}\quad \beta_e>1
\end{equation}
The latter condition, with $\beta_e=2\mu_0NT_e/ B^2$ is the requirement that $r_{ce}>\lambda_e$ which must be satisfied for spontaneous reconnection to occur. It corresponds to the condition that the half-width of the current filaments is of the order of the electron inertial scale \citep{treumann2015a}.  This condition holds allover in the solar wind. Hence, the above formula provides an estimate for the equivalent anomalous collision frequency in spontaneous reconnection in the solar wind turbulence. 

The dependence on the energy input is reasonable as this is the energy which enters into the formation of the electron-scale current filaments when they undergo spontaneous reconnection. On the other hand, measuring the turbulent dissipation rate provides an estimate of the energy injection rate into the turbulence. The last formula can be written 
\begin{equation}
\epsilon_\mathit{in}\sim \nu_\mathit{an}^3\lambda_e^2
\end{equation}
which expresses the energy per time and mass injected into the inertial range in MHD turbulence through the anomalous collision frequency and the electron-inertial length.

If we now adopt the simple argument that the mean-free path of electrons $\lambda_\mathit{mfp}\sim v_e/\nu_\mathit{an}$ is the ratio of the electron thermal speed $v_e$ and the (anomalous) collision frequency, we can provide an estimate of the latter during reconnection. Simulations have shown that the electron temperature in a single electron exhaust can go up by a factor of 100  while the one-sided length of the exhaust is of the order of roughly $100\, \lambda_e$ \citep[cf., e.g.,][]{le2015}. Hence, with $\lambda_\mathit{mfp}\approx\, 100\,\lambda_e$ the anomalous collision frequency that is produced in spontaneous collisionless reconnection by the electron exhaust can be estimated to be of the order of $\nu_\mathit{an} \sim 0.1v_\mathit{e}/\lambda_e$. With the Debye-length $\lambda_D=v_e/\omega_e$ this can be written as 
\begin{equation}
\nu_\mathit{an}/\omega_e\sim 0.1 (\lambda_D/\lambda_e) =\,0.14\  (T_e/m_ec^2)^\frac{1}{2}\, \lesssim\, 10^{-3} 
\end{equation}
This estimate holds in the solar wind with $T_e\approx 100$ eV, the value at which the observations of \citet{brown2015} were performed. The solar wind density was $N\approx 10^7$ m$^{-3}$, yielding $\omega_e\approx 220$ kHz. This gives quite a substantial anomalous collision rate of 
\begin{equation}
\nu_\mathit{an}\lesssim 200\quad\mathrm{Hz}
\end{equation}
in the solar wind caused by spontaneous reconnection in the thin turbulent current filaments that have been generated in the inertial range of turbulence. This collision rate would be high enough for dissipating the energy inflow. 

It is interesting to compare it with another independent estimate  \citep{treumann2014}\footnote{In Eq. (22) of that paper a typo occurred. The exponent of the diffusion coefficient should be 0.17 ($\sim 1/6$) instead of 1.17.} that has been based on numerical 2d-simulations of reconnection. There it was found that the anomalous collision rate generated by collisionless reconnection amounted to $\nu_\mathit{an} \sim\, 0.03\ \omega_{ce}$, where $\omega_{ce}$ is the electron cyclotron frequency of the ambient magnetic field -- note that this is in contrast to the above use of the electron plasma frequency $\omega_e$! For the above conditions this yields $\nu_\mathit{an}\approx\, 53$\,Hz in reasonable agreement with the above rough estimate. 

The electron inertial length is of the order of $\lambda_e \approx1.5$ km. These numbers may be used in an estimate of the injected energy flux which then yields that energy of a rate  
\begin{equation}
1\ \mathrm{eV}\,\mathrm{s}^{-1}\ \lesssim\ \textstyle{\frac{1}{2}}m_e\epsilon_\mathit{in}\  \lesssim\ 50\  \mathrm{eV}\,\mathrm{s}^{-1}
\end{equation}
has been injected into the turbulence in order to become dissipated by reconnection under the assumed stationary conditions. This is the power of the turbulent injection. Little is known about the fate of this energy flux. A fraction ends up generating the electron halo in the electron distribution, some fraction contributes to heating, and some small amount feeds the generation of genuine plasma waves like Langmuir and ion-acoustic waves.

\section{Fate of injected energy}
Observations have so far just concentrated on the identification of the inertial range, determination of the spectral slope, confirmation of Kolmogorov's theory, and inference on the dissipation range. The physically interesting question concerns the fate of the injected energy. That, under stationary conditions, this energy must be dissipated in some way, is quite natural. By what process this dissipation takes place is another question. It implies that under stationary conditions dissipation must not simply balance the injection, it must also consist of a number of different processes which include transport of energy away from the turbulence. If dissipation, as is naively assumed, just heats the medium continuously, the plasma will readily be heated up to unreasonably high energies until becoming glowing. Since in the solar wind this is unreasonable and is not observed, the question of balancing energy injection is a sensible one. 

In this respect spontaneous collisionless reconnection in narrow current layers or current filaments as the ultimate dissipation mechanism of magnetic turbulence has a number of advantages over any other proposed dissipation mechanism. Heating is but one, and presumably not even the most important part of this process. The main effect of spontaneous reconnection is the acceleration of particles to high energies \citep[for a comprehensive review cf., e.g.,][]{treumann2015b}. 
Electrons become accelerated in the electron exhaust. They subsequently experience further acceleration by interaction with neighbouring reconnection sites and plasmoids. They end up in a process which can be described as Fermi. The solar wind, on the other hand, is a weakly magnetised medium with moderate energy injection. Acceleration of electrons just causes the formation of wings on the electron distribution, the solar wind electron-halo distribution.

However, the solar wind is by no means stationary. It is continuously flowing and expanding. Hence, assuming that the above constant energy injection into the solar wind would become transferred solely into heat, one may ask whether and how it can be balanced by convection. 

At the moderately high solar wind speeds \citep[in the case of][they amounted to super-magnetosonic velocities $V_{sw}\approx\, 600$ km\,s$^{-1}$]{brown2015} one expects that any heating would be balanced by isentropic (adiabatic) expansion. Since expansion is just radial, this implies that the solar wind temperature between two radial distances $R_2,R_1$ adiabatically decreases with volume. 

Let us assume that the heating is balanced solely by adiabatic cooling (holding entropy $S$ constant). Applying the ordinary adiabatic law with index $\gamma=5/3$, one has for the temperatures $T_2/T_1=(R_1/R_2)^2$. The observations have been performed at $R_1\approx 1$\,AU and $T_1=T_e\approx 100$\,eV. At the above speed, the solar wind at $R_1$ is supplied $\Delta W =\frac{3}{2}\Delta T\, \approx\, 50$ eV of energy during 1 s, i.e. over a $\Delta R\, \approx\, 600$ km distance of flight. The isentropic cooling in temperature over this same distance amounts to
\begin{equation}
\Delta T \,\equiv\, T_2-T_1\,=\, -2\,T_1\Delta R/R_1\, \approx\ 8\times10^{-4}\quad\mathrm{eV}
\end{equation}

This number seems to be insufficient to balance the above heating. However, in obtaining the cooling we have considered the increase of the entire solar wind volume, while the dissipation regions of the involved current filaments just occupy a small fraction of the volume. Hence, a volume-filling factor $f_{cs}$ of current sheets \citep{treumann2015b} must be taken into account when balancing the two numbers:
\begin{equation}
f_{cs}\,\approx\, 3\times 8\times10^{-4}/2\times(1-50)\ \sim\ 10^{-4}\,-\,10^{-5}
\end{equation}

The expectation of balancing the energy supply solely by expansion is of course highly exaggerated. The above volume filling factor can probably be reduced by another order of magnitude or so. 

As mentioned above, most of the  energy dissipated in spontaneous reconnection does not go into heating. It is consumed by the acceleration of a comparably small number of electrons to comparably high energies. These electrons form the wings of the solar wind electron distribution, the electron halo. Non-relativistic  simulations suggest a factor of 10 to 100 in electron energy \citep[cf.,][]{treumann2015b}. The halo population transports most of the turbulent energy away in the form of heat flux and dissipates it in other ways like the generation of kinetic plasma waves. It is just the remainder that may be balanced by adiabatic cooling.

\section{Thermodynamic interpretation}
Kolmogorov theory provides information about the overall entropy spectrum d$S(k)$ (in non-dimensional units, i.e. normalised to the Boltzmann constant and mass) expressed through the energy d$W$ per mass. Since in the inertial range no work is done on the volume with d$V=0$, this is given by  the first law:
\begin{equation}
\theta\mathrm{d}S(k)=\mathrm{d}W(k)=W(k)\mathrm{d}k
\end{equation}
with $\theta=T/m$ the temperature per mass. Inserting for $W(k)$ from Eq. (\ref{eq-wk}) and neglecting the small additional exponent $\delta$ yields after performing the integration over $k$ in the intervall of the inertial range
\begin{equation}
S(k)={\frac{3}{2}}\frac{C_K}{\theta}\left(\frac{\epsilon_{in}}{k_\mathit{in}}\right)^{2/3}\left[1-\left(\frac{k_\mathit{in}}{k}\right)^{\!\!\!\!2/3}\right],
\quad k\leq k_d
\end{equation}
This is the spectrum of the microcanonical entropy which belongs to the Kolmogorov inertial range. 
From statistical mechanics we know that the microcanonical entropy is the logarithm of the probability $w_k$ for the state $k$ to be realised. Hence we have
\begin{equation}\label{eq-prob}
w_k=Z^{-1}\ \exp\left[\xi -\frac{3C_K}{2\theta} \left(\frac{\epsilon_\mathit{in}}{k}\right)^{2/3}\right]
\end{equation}
with $Z(\theta,\epsilon_{in})=\int_{k_{in}}^{k_d}\mathrm{d}k\, \exp[S(k)]$ the partition function as the integral over the entire inertial range. The latter integral is not easy to perform. We may assume that $k_\mathit{in}$ is small and the dissipation wave number $k_d$ large, in which case the term in the brackets can be simplified, and the limits of the integration become 0 and $\infty$. On the other hand, we may also define $\eta\equiv C_K\epsilon_\mathit{in}^{2/3}/\theta, \zeta_{d,\mathit{in}}\equiv\eta(2\pi/k_{d,\mathit{in}})^{3/2}$ to find (after substantial amount of algebra) the explicit partition function
\begin{eqnarray}
{Z(\theta,\eta)}&=&4(\pi\eta)^\frac{3}{2}\mathrm{e}^{\xi}\left[\mathrm{erf}\sqrt{\zeta_\mathit{d}}\ -\ \mathrm{erf}\sqrt{\zeta_\mathit{in}} \right. \ +\cr
&-&   \left.                                                                                                                                                                                                                                                                                                                                                                                                    \frac{\mathrm{e}^{-\zeta_d}}{(2\zeta_d)^{3/2}}(1-2\zeta_d)+\frac{\mathrm{e}^{-\zeta_\mathit{in}}}{(2\zeta_\mathit{in})^{3/2}}(1-2\zeta_\mathit{in})\right]
\end{eqnarray}
with $\xi\equiv \frac{3}{2}\zeta_\mathit{in}/(2\pi)^\frac{3}{2}$. Since $\zeta_{in}\gg\zeta_d$ the main contribution to the partition function comes from  the region close to the dissipation range, which shows that the dissipation is essential for the formation of the spectrum, the assumption made explicitely by \citet{kolmogorov1941a}. This last expression can in principle be used to obtain the free energy in the inertial range (now in energy units)
\begin{equation}
F= -T \log Z(\theta,\eta)
\end{equation}
which links the Kolmogorov-range turbulence to statistical mechanics. The free energy depends on the injection rate $\epsilon_\mathit{in}$, injection and dissipation wave numbers $k_d, k_\mathit{in}$, and on the temperature $T$. 

The above probability Eq. (\ref{eq-prob}) ist actually the probability spectrum belonging to the Kolmogorov inertial range. It is of the form of a fractal probability spectrum 
\begin{equation}
w_k=A\,\exp(-\epsilon_\mathit{in}/k)^p
\end{equation}
with fractional power $p=2/3$ and
\begin{equation}
A=Z^{-1}\exp\left[\frac{3}{2}\frac{C_K}{\theta}\left(\frac{\epsilon_\mathit{in}}{k_\mathit{in}}\right)^{\!\!\!\frac{2}{3}}\right]
\end{equation}
Its inverse Fourier transformation back into configuration space cannot be given in closed form. The turbulent inertial range is neither stochastic nor is it subdiffusive. It indicates superdiffusive and thus chaotic interactions. Thie probability  can  be used to calculate the most probable inertial-range expectation value $\langle \mathcal{O}\rangle=\int\mathcal{O}_kw_k\mathrm{d}k$ of any desired observable $\mathcal{O}$. 

The inertial range is not undergoing any violent dissipation. However, each of the turbulent eddies at any fixed $k$ must pick up energy from the eddy at the adjacent smaller wavenumber $k-\Delta k$ and transport it up to the next higher wavenumber $k+\Delta k$. It is thus in contact on the one side with an energy reservoir of higher energy and on the other side with another reservoir of lower energy to which it transports the energy and then returns to its initial state in order to pick up the next amount. 

This is a Carnot cycle which the eddy performs in the magnetohydrodynamic fluid. Though the process for the particular eddy is about reversible, it is irreversible when considering the eddy in contact with its environment, because the energy flow is in one and only one direction. It is extracted from the adjacent larger-scale eddies and transferred to the adjacent smaller-scale eddies. There is no compensating reverse flow, and hence the transport of energy implies that the overall system is subject to an irreversible process that is not isentropic even though we have not yet reached into the dissipative range where the energy will ultimately become destroyed. The amount of  energy we calculated above is therefore equal, by the first law of thermodynamics, to the amount of entropy $\Delta S$ that the inertial range contributes under stationary conditions and at constant volume:
\begin{equation}
\theta\mathrm{d}\Delta S(k)\ =\ \mathrm{d}\Delta W(k)\ =\ \delta W(k)\mathrm{d}k
\end{equation}
where d$\Delta S(k)$ is the differential of the additional entropy generated due to the slope mismatch $\delta$. For the MHD turbulence under consideration $W(k)$ is again given in Eq. (\ref{eq-wk}). Dividing by the temperature $T$ and integrating with respect to $k$ yields the entropy increase in the inertial range. An estimate (now switched back to energy units) is obtained when using the  average energy density $\langle\Delta W_B\rangle$ in solar wind magnetic fluctuations given in Eq. (\ref{wb}): 
\begin{equation}
\Delta S\ =\  \frac{1}{T}\langle W_B\rangle\ \approx\ 4\times10^{-3}\frac{W_B}{T}\ \approx\ 4\times10^{-3}\beta_{e}^{'-1} 
\end{equation}
where $\beta_e'=2\mu_0NT_e/\langle|\delta B^2|\rangle$ is the plasma beta with respect to the average inertial range wave power. If we use $T_e\approx 100$ eV as suggested by the observations, the stationary increase in the (dimensionless) entropy produced in magnetohydrodynamic turbulence in the Kolmogorov range becomes 
\begin{equation}
\Delta S\ \approx\ 4\times 10^{-5}
\end{equation}

We emphasise that this number applies solely to the extra entropy generated in the inertial range of the spectrum. The main dissipative entropy production by the solar wind (or any MHD) turbulence takes of course place in the dissipation range where the turbulent energy is ultimately transformed into other forms like heating, particle acceleration, topological change of the magnetic field, destruction of turbulent vortices and currents and generation of kinetic plasma turbulence. These are complex processes that cannot simply be inferred from investigation of the spectral slope of turbulence. 

The difference between the numerical simulation and the ideal inertial theory thus provides an estimate of the entropy that is contributed inside the inertial range in order to keep the inertial range eddies going. This process consumes some work which therefore is taken out of the turbulent energy. Entropy is increased by the mere action of energy transport down the inertial range from large to short scales under stationary conditions. 

The stationary state of turbulence can be kept alive only by continuous injection of energy at large scales. This necessarily leads to irreversible generation of entropy already in the cascade. Any stationary turbulence theory belongs to the domain of thermodynamics respectively statistical mechanics. So far it does neither refer to temperature nor to partition function. This poses an open gap  in stationary turbulence theory. 


\section{Conclusions}
\vspace{-0.1cm}
In the present note we compared Kolmogorov theory of magnetic turbulence with high-resolution observations \citep{brown2015} of magnetic turbulence in the solar wind and with high resolution magnetohydrodynamic simulations \citep{bereznyak2014}. In the observations we identified a subrange in the spectrum at higher wave numbers $k$ which had the exact slope of the spectra obtained in the simulations. 

The significant though small difference in the slopes we took as an thermodynamic indication for the maintenance of the inertial range spectrum. This provided an estimate of the energy that is required to maintain a stationary spectrum. Naturally this is only a small fraction of the injected energy flux, of the order of per mille only. Its significance is to bring turbulence theory into accord with thermodynamics. It also provides an estimate of the entropy production in Kolmogorov turbulence. 

These results enabled us to estimate the anomalous collision frequency in the dissipation range based on spontaneous reconnection and with its help the energy injection rate into the solar wind turbulence. When discussing the fate of the injected energy we found that isentropic expansion of the solar wind may balance the heating by dissipation of the turbulent energy. This requires considering a large number of reconnecting current filaments at small volume-filling factor. 

However, it is unlikely that the energy input is balanced by isentropic expansion and by no means by expansion alone. Dissipation in spontaneous reconnection generates high energy particles in the first place. These occupy the solar-wind electron halo distribution. Solar wind turbulence and its dissipation by spontaneous reconnection is thus identified as being its cause.    
\vspace{-0.15cm}
%
%





\end{document}